 \documentclass[nohyper,12pt,letterpaper]{JHEP3}
 \usepackage{epsfig}
 
 \def\bR{{\mathbb R}}

\title{Semiclassical Unimodular Gravity}

\author{Bartomeu Fiol and Jaume Garriga\\

 Departament de F{\'\i}sica Fonamental i \\Institut de Ci{\`e}ncies del Cosmos, 

Universitat de Barcelona,\\

Mart{\'\i}\ i Franqu{\`e}s 1, 08028 Barcelona, Spain\\

\email{bfiol@ub.edu, jaume.garriga@ub.edu}}

\abstract{Classically, unimodular gravity is known to be equivalent to General Relativity (GR),
except for the fact that the effective cosmological constant $\Lambda$ has the status of an integration constant. 
Here, we explore various formulations of unimodular gravity beyond the classical limit.
We first consider the non-generally covariant action formulation in which the determinant
of the metric is held fixed to unity. We argue that the corresponding quantum theory is also equivalent to 
General Relativity for localized 
perturbative processes which take place in generic backgrounds of infinite volume (such as asymptotically flat 
spacetimes). Next, using the 
same action, we calculate semiclassical non-perturbative quantities, which we expect will be dominated by Euclidean instanton solutions. 
We derive the entropy/area ratio for cosmological and black hole horizons, 
finding agreement with GR for solutions in backgrounds of infinite volume, but disagreement for backgrounds with 
finite volume.
In deriving the above results, the path integral is taken over histories with fixed 4-volume. We point out that 
the results are different if we allow the 4-volume of the different histories to vary over a continuum range. 
In this "generalized" version of unimodular gravity, one recovers the full set of Einstein's equations in the 
classical limit, including the trace,
so $\Lambda$ is no longer an integration constant. Finally, we consider the generally covariant theory due to Henneaux and Teitelboim, which 
is classically equivalent 
to unimodular gravity. In this case, the standard semiclassical GR results are recovered provided that the boundary term in the Euclidean action 
is chosen appropriately.}

\begin{document}

\section{Introduction and conclusions}

Despite the many successes of General Relativity, unresolved issues like the cosmological constant problem, 
or the problem of time, have prompted physicists to formulate alternative theories of gravity, in the hope 
that they might retain the good features of General Relativity and be able to tackle from a new angle the 
problems mentioned before. {\it Unimodular gravity} is such an alternative theory. There, we only 
consider metrics whose determinant is fixed to be (minus) one, so unlike in General Relativity not every 
diffeomorphism is allowed. The only diffeomorphisms allowed are those that satisfy 
$g_{\mu \nu}\delta g^{\mu \nu}=0$. 
One of the possible actions for unimodular gravity (modulo a boundary term to be added below) can be written as
\begin{equation}
I=\frac{1}{16\pi G}\int d^4 x (R-2L)
\label{unimod}
\end{equation}
where the metric has unit determinant. Note that we added a constant $L$; it does not appear in the equations of motion, 
so we refrain from identifying it with the cosmological constant of ordinary General Relativity. However it might 
appear in semiclassical results, in the same way as $\theta_{QCD}$ does, even if it does not appear in the equations 
of motion. 

The variation of the action yields
$$
g^{ab}\delta R_{ab}+\delta g^{ab}R_{ab}=0
$$
The first sumand is the same total divergence as in GR, so the equations of motion are derived from
$\delta g^{ab}R_{ab}=0$. Now, however, the components of $\delta g^{ab}$ are not independent, they 
satisfy $g^{ab}\delta g_{ab}=0$, so we can't conclude $R_{ab}=0$. We can only conclude that its 
traceless part is zero, so the resulting equations of motion are 
$$
R_{ab}-\frac{1}{4}Rg_{ab}=0
$$
note the $1/4$ instead of the ordinary $1/2$ (in D dimensions, one gets 1/D) \footnote{These equations of motion were actually 
written down by Einstein himself \cite{einstein}, during the early days of General Relativity.}.  If we take the trace we 
get $0=0$, so we can no longer conclude that $R_{ab}=0$. However, using the Bianchi identity 
we can conclude that $R$ must be a constant, so we can rewrite the equation of motion as
$$
R_{ab}-\frac{1}{2}Rg_{ab}=\Lambda g_{ab}
$$
We see that in this formulation the cosmological constant appears an as integration constant in the equations of motion, rather 
than as a parameter in the Lagrangian: even if we allow for $L\neq 0$ in (\ref{unimod}), $\Lambda$ and $L$ have no reason to be related. De Sitter and Anti de Sitter spaces are now solutions of the vacuum equations. Over the years, this theory has been considered 
by a number of people, chiefly due to the different perspective it brings to the cosmological constant problem \cite{Weinberg:1988cp}.

In General Relativity the Hilbert-Einstein action 
is complemented by a boundary term \cite{Gibbons:1976ue}. 
The derivation of the Gibbons-Hawking boundary term in unimodular gravity \cite{diego} yields the same result as in ordinary 
General Relativity (see e.g.  appendix E of \cite{wald} for details on the derivation), so the action (\ref{unimod}) is complemented by a term
\begin{equation}
I_{bdy}=\frac{1}{8\pi G}\int_{\partial Y}d^3x \; \sqrt{-h}K.
\label{ibdy}
\end{equation}

Classically, unimodular gravity yields the same predictions as ordinary General Relativity, and so it satisfies the common phenomenological 
tests of this last theory, but in principle quantum effects could be used to discriminate between the two theories 
\cite{Alvarez:2005iy}. The main purpose of this note is to explore this possibility. 

In Section 2, we argue that for localized perturbative processes 
which take place in an asymptotically flat space (or more generically, in any background spacetime of infinite volume) 
the quantum theory based on the action (\ref{unimod}) is equivalent to GR. This conclusion seems to exclude the possiblity of
phenomenologically testable differences with GR in the perturbative domain, so we go on to consider semiclassical non-perturbative processes
in various formulations of unimodular gravity.

We begin by considering the theory based on a path integral formulation of action (\ref{unimod}) plus the boundary term (\ref{ibdy}).
An important ingredient which must be specified in this discussion is what histories are we supposed to be summing over. In GR, all histories satisfying 
given boundary conditions are included. In unimodular gravity, we seem to have two possibilities. The first one is to keep the coordinate 4-volume 
(and hence the physical 4-volume) fixed in the sum over histories. The second possibility is to allow for histories
with different four-volumes. Note that the distinction does not exist in GR, since a fixed coordinate volume does not restrict the physical volume.

In Section 3, we explore the case where the 4-volume is held fixed in the sum over histories.
When evaluating the action of a Euclidean instanton in backgrounds with infinite volume, one obtains an infinite result that has to be regularized. 
The standard procedure is to subtract the action of a reference background solution. In the absence of matter, any solution of Einstein's 
equation has constant $R$, $R=4\Lambda$, so the integrand in the bulk term in the action is constant 
and gets out of the integral
\begin{equation}
I=\frac{1}{16\pi G}(4\Lambda-2\Lambda)\int d^4 x \sqrt{-g}.
\end{equation}
The last factor is the 4-volume, and it should not depend on whether we use a unimodular metric or not. 
However, the factor in front of it does depend on the action used: in unimodular gravity the factor 
$(4\Lambda-2\Lambda)$ gets replaced by $(4\Lambda-2L)$. 
If we were to take the instantonic solution and the reference background to have the same value of $\Lambda$ (but different total four-volumes),
this would result in the unimodular action being $(2-L/\Lambda)$ times the GR action. By the standard identification of the Euclidean action
with the logarithm of the partition function, this would affect the evaluation of the energy and entropy of the solution. For instance, 
for black hole solutions in AdS, and for any $L\neq\Lambda$, the energy $E$ calculated from the partition function would
not agree with the mass parameter $M$ of the black hole solution, leading to inconsistencies with the first law of thermodynamics. 

Instead, here we propose that the subtraction should be made by keeping the 4-volume of the instanton and the reference background to be 
the same. The two solutions should of course share the same geometry at the regulating boundary, which we take to be a two-sphere of
large intrinsic radius $R$, times the thermal circle. The 4-volumes corresponding to the two geometries (with and without the black hole) 
can be kept to be the same provided that we adjust the corresponding values of the integration constants $\Lambda$ and $\Lambda'$. 
In the limit when we remove the regulator ($R\to \infty$), $\Lambda$ and $\Lambda'$ coincide, and the regularized action 
reproduces that of ordinary GR for the Schwarzschild-AntideSitter black hole.

So far, we discussed instantons in backgrounds with infinite volume. When we turn to instantons with finite volume, corresponding to positive $\Lambda$,
the outcome is drastically different. Here there is no need to regularize the action of the instanton, since it is already finite, and the results 
disagree with the ones obtained in GR. It is not clear to us whether these results admit a sensible thermodynamical interpretation, but on the other hand we haven't found a sharp contradiction 
with the laws of thermodynamics either. Some of the complications that prevent us from reaching a definite conclusion are that in de Sitter there is 
no well defined notion of energy\footnote{Of course, an approximate notion of energy exists on scales much smaller than the de Sitter radius, but 
for small black holes the Schwarzschild-de Sitter solutions do not have a smooth Euclidean section, since the two horizons are at different temperatures.}. 
Clearly, this issue deserves further study.

Next, in Section 4, we discuss the case where the 4-volume is allowed to vary in a continuum range in the sum over histories.
In this case, we have some freedom in weighing the contributions from histories with different values of the 4-volume. This leads to
a family of theories which generalizes (\ref{unimod}) by the addition of a single global degree of freedom, 
corresponding to the constant value of the determinant of the metric. These theories agree with GR in the semiclassical limit. In particular the 
effective cosmological constant is no longer an integration constant but it is completely determined by the Lagrangian. Nonetheless, 
the theories are not generally covariant, and may differ from GR away from the semiclassical limit.

Finally, in Section 5  we consider the generally covariant theory introduced by Henneaux and Teitelboim \cite{Henneaux}, which is classically 
equivalent to (\ref{unimod}). In this case, we show that there is a choice of the boundary term for which the standard semiclassical results of GR 
are recovered, while the effective cosmological constant remains an integration constant.

The rest of the note is structured as follows. Section 2 is devoted to the quantum equivalence of GR and (\ref{unimod}) for localized perturbative
processes taking place in spacetimes of infinite volume. In Section 3 we evaluate the action (\ref{unimod}) for the instantons obtained from the 
Euclidean continuation of various vacuum solutions. For the Schwarzschild black hole in asymptotically Anti de Sitter and asymptotically flat 
spaces, we show that our regularization procedure yields the same finite answer as the respective GR computations. On the other hand, 
for the de Sitter solution, the result we obtain disagrees with the one obtained in GR.
Section 4 discusses the generalized unimodular theories, where one integrates over the constant value of the determinant with an 
arbitrary weight function. Section 5 is devoted to the generally covariant formulation of Henneaux and Teitelboim. 

\section{Perturbative equivalence of GR and unimodular gravity}

Let us consider the generating functional 
$$ 
Z[J] = \int Dg_{\mu\nu} D\psi e^{iI[g_{\mu\nu},\psi,J]}. 
$$ 
Here $I$ denotes the action of GR, including boundary terms, plus the action for matter fields $\psi$, and $J$ 
denotes external sources. The action is invariant under diffeomorphisms, generated by arbitrary vector 
fields $\xi^{\mu}(x)$, and so is the measure $Dg_{\mu\nu}$. 

A general vector field can be decomposed into transverse and longitudinal part $\xi=\xi_t +\xi_l$. 
The transverse diffeomorphisms form a subgroup . On the other hand, since 
\begin{equation} 
\delta \sqrt{g}= {1\over 2} \sqrt{g}\ \nabla_\mu \xi^{\mu},\label{dg} 
\end{equation} 
it is clear that we can bring the determinant 
of the metric to unity in the neighborhood of any given point by using longitudinal diffeomorphisms 
\begin{equation} 
g(x) \equiv |\det g_{\mu\nu}(x)| = 1. 
\end{equation} 
However, if we demand 
that the gauge transformation vanishes at infinity (or at the boundary of a prescribed portion 
of spacetime)\footnote{We thank Takahiro Tanaka for drawing to our attention the convenience of this requirement.}, then the average value of the determinant 
\begin{equation} 
a[g] \equiv {\int_{\cal M} \sqrt{g} d^4 x \over \int_{\cal M} d^4 x},\label{av} 
\end{equation} 
cannot change. Indeed, the spacetime volume 
\begin{equation} 
V=\int \sqrt{g} d^4 x, 
\end{equation} 
is gauge invariant when we require the vanishing of gauge transformations at the boundary
: 
\begin{equation} 
\delta V = {1\over 2} \int_{\cal M}  \sqrt{g}\ (\nabla_\mu \xi^{\mu}) d^4 x= \int_{\partial{\cal M}} \xi^\mu d\Sigma_\mu =0. 
\end{equation} 
With this in mind, let us partially fix the gauge in the path integral by using the Fadeev-Popov trick.  The
motivation for partially fixing the gauge is that the unimodular theory is invariant only under the transverse
subgroup of diffeomorphisms, and this will highlight the relation between both theories. We start with
the identity
\begin{equation}
1=\Delta[g_{\mu\nu}] \int D(\partial_{\mu}\xi^{\mu}) \delta\left({\sqrt{g}\over a[g]}-1\right),\label{identity}
\end{equation}
where we introduce the functional determinant
$$
\Delta[g_{\mu\nu}]=\left| {D\left({\sqrt{g}\over a[g]}\right) \over D (\partial_\mu \xi^\mu)}\right|_{\sqrt{g}= a[g]}.
$$
Using (\ref{dg}), it is straightforward to check that $\Delta$ is a constant, 
completely independent of the metric. Hence, it is gauge invariant. Inserting the left hand side of (\ref{identity})
into the path integral, we have 
\begin{equation} 
Z[J] \propto \int Dg_{\mu\nu} D\psi \ \delta\left({\sqrt{g}\over a[g]}-1\right) 
e^{iI[g_{\mu\nu},\psi,J]}, \label{pgf} 
\end{equation} 
where we have factored out the constant $\Delta$, and we have also used the gauge invariance of the action and of the measure
in order to factor out the infinite ``volume" $\int D(\partial_{\mu}\xi^{\mu})$.
The constraint can be exponentiated by introducing a Lagrange multiplier $\lambda(x)$: 
\begin{equation} 
Z[J] \propto \int Dg_{\mu\nu} D\psi D\lambda\ 
e^{iI[g_{\mu\nu},\psi,J]+i \int d^4 x\ \lambda \left({\sqrt{g}\over a[g]}-1\right)}. 
\end{equation} 
This should still be the generating functional for GR, in spite of the fact that the action in the exponent 
is only invariant under transverse diffeomorphisms (due to the partial gauge fixing). Let us check that the 
classical equations are precisely the same as in GR. Unrestricted variation with respect to the metric gives 
\begin{equation} 
G_{\mu\nu} +\Lambda g_{\mu\nu} - 8\pi G T_{\mu\nu} = 8\pi G 
\left( {\lambda(x) \over a[g]} - { \int_{\cal M} \lambda \sqrt{g} d^4 x \over a^2[g] \int_{\cal M} d^4 x}\right)g_{\mu\nu} . \label{eel} 
 \end{equation} 
Here, $T_{\mu\nu}$ is the energy momentum tensor of matter fields, and a cosmological term which may be present in the 
gravitational action has been displayed 
explicitly. Variation with respect to $\psi$ yields the matter field equations, which in turn imply 
$$ 
T^{\mu\nu}_{\ \ ;\nu} =0. 
$$ 
Taking the covariant divergence of Eq. (\ref{eel}) we have $\lambda(x) = const.$, and from this, it follows that the 
right hand side of Eq. (\ref{eel}) is precisely zero. Hence, 
\begin{equation} 
G_{\mu\nu} +\Lambda g_{\mu\nu} = 8\pi G T_{\mu\nu}, 
\end{equation} 
as it should be the case in GR. Note that, unlike the case of unimodular gravity, here we do not have the freedom of adding 
an arbitrary constant to the cosmological term. Finally, variation with respect to $\lambda$ fixes the gauge in such a way 
that the determinant of the metric is a constant. 

Once we have cast the generating functional in the form (\ref{pgf}), a corollary follows: 
\begin{itemize} 
\item For any localized perturbative process which takes place in an asymptotically flat spacetime 
(or in any spacetime of infinite volume), the predictions of GR coincide with those of unimodular gravity. 
\end{itemize} 

Indeed, if the process is localized in Minkowski space, we only need to sum over histories which differ significantly from Minkowski in 
a finite spacetime region. Hence, we are justified in requiring that perturbations fall 
off to zero sufficiently fast at large distances from the process we are studying. For those histories, the average of the 
square root of the determinant of the metric will be equal to 1 in the limit of infinite spacetime volume, and we can set $a=1$ 
as a boundary condition in the path integral.  With this boundary condition, Eq. (\ref{pgf}) coincides with the 
generating functional of unimodular gravity. 

The same argument goes through for any background of infinite volume, not necessarily Minkowski. We start by casting the background in 
unimodular form, and ask that the perturbations we are integrating over vanish sufficiently fast away from the localized process.

\section{Semiclassical calculations at fixed 4-volume}

In this section we consider the evaluation of the action (\ref{unimod}) plus (\ref{ibdy}) for various Euclidean solutions.

We start by considering solutions in backgrounds with infinite volume, namely the Euclidean versions of the Schwarzschild black hole solution in 
asymptotically flat space and the Schwarzchild-Anti de Sitter black hole solution. 
The evaluation of the action for these solutions is well-known in GR \cite{Gibbons:1976ue, Hawking:1982dh}, 
but here we face the delicate issue of which reference background should we use for subtraction. As we argued in the Introduction,
in the present case the instanton solution and the background should have the same coordinate 4-volume. 
With this prescription, we find that the regularized action coincides with the one found in ordinary GR in the limit of infinite 
volume (although the result arises in a somewhat non-trivial manner). This leads to the standard thermodynamical properties.

We then turn to a solution with finite 4-volume, the deSitter solution. Here there is no need or regularization, as the action is already finite. 
However, as we will see, we get a value for the action different from the one in General Relativity. 
Furthermore, the resulting entropy/area relation not only differs from the one in GR (i.e. 1/4), but it is no longer universal, 
unless we 'lock'  the value of $L$ to that of $\Lambda$, $L=\Lambda$. For instance, for $L=0$, we get an entropy/area factor of 1/2 for any  
$\Lambda > 0$.

\subsection{Instantons in Minkowski/Anti de Sitter.}
Our starting point is the Schwarzschild solution in asymptotically flat space ($\Lambda=0$),
$$
ds^2=-\left(1-\frac{2MG}{r}\right)dt^2+\frac{1}{1-\frac{2MG}{r}}dr^2+r^2d\Omega ^2
$$
The change of coordinates that brings this metric to unimodular form can be found in the original Schwarzschild paper 
\cite{schwarz}\footnote{We would like to thank Roberto Emparan for informing us that in the derivation of the solution 
that carries his name, Schwarzschild actually imposed the unimodular condition on the metric.}
\begin{equation}
\rho=\frac{r^3}{3} \hspace{1cm} x=-\hbox{cos }\theta
\label{changesc}
\end{equation}
gives the unimodular metric
\begin{equation}
ds^2=-\left(1-\frac{2MG}{r}\right)dt^2+\frac{1}{1-\frac{2MG}{r}}\frac{1}{r^4}d\rho^2+
r^2\left(\frac{1}{1-x^2}dx^2+(1-x^2)d\phi^2\right)
\label{unimodbh}
\end{equation}
This metric is not well defined for $x=-1,1$, i.e. the North and South pole of the $S^2$, 
but we think this is not worrisome. Clearly those are coordinate singularities, and one could 
use more than one patch to cover the sphere, and that won't affect the evaluation of the action. 
>From now on, we denote by $d\bar \Omega_2^2$ the unimodular metric of $S^2$, as in eq. (\ref{unimodbh}).

We go to Euclidean signature by defining an imaginary time $\tau =it$.  Now we come to the 
issue of the temperature of this black hole in unimodular gravity, or equivalently what is the 
period of $\tau$. In General Relativity, one can find this temperature by different arguments. 
The first one is the classical computation of Hawking \cite{Hawking:1974sw}. Since it only involves the metric, and not the action, 
an equivalent computation in unimodular gravity ought to yield the same result.\footnote{Pragmatically, 
if for some subtle reason that escapes us, it would produce a different temperature,  this would indeed 
prove that the two theories differ (even in asymptotically flat spacetimes).} 

Another way to derive the periodicity of $\tau$, as argued in \cite{Gibbons:1976ue}, 
is to write down the solution in Kruskal coordinates, and require that the metric has no singularities. 
Also, this shows the existence of a section of the complex metric without singularities. We couldn't 
find some Kruskal-type change of coordinates that keeps the metric unimodular, but we will now show that very 
near the horizon $r=2MG$, one can write a unimodular metric that is the product of  $\bR^2 \times S^2$, 
provided one takes $\tau$ to be periodic, with the same period as in General Relativity. The change of 
coordinates 
$$
r-2MG=\frac{x^2}{ 4(2MG)^3}
$$
very near the horizon, gives the approximate metric
$$
ds^2\approx \frac{1}{(2MG)^2}\left(x^2d(\frac{\tau}{4MG})^2+dx^2\right)+(2MG)^2d\bar \Omega_2^2
$$
the first factor is the Euclidean version of Rindler space, which will be $\bR^2$ if we assign periodicity $8\pi MG$ to 
$\tau$. Then $\bR^2$ can be brought to Cartesian coordinates and the full metric is unimodular.

We then proceed granting that there is a 
full change of coordinates bringing the metric to unimodular form - of which we only 
presented the version near the horizon - that shows the existence of a section that avoids the 
singularity at the origin, and that this change of coordinates requires that the periodicity of $\tau$ 
in the unimodular case has to be taken the same as in GR.

The discussion above is easily generalized to the Schwarzschild-Anti de Sitter metric. 
The metric of the covering space of Anti deSitter space in static coordinates is ($b^2=-3/\Lambda$
, with $\Lambda<0$)
$$
ds^2=-Vdt^2+V^{-1}dr^2+r^2d\Omega_2^2
$$
with 
$$
V=1+\frac{r^2}{b^2}
$$
The Schwarzschild-Anti de Sitter metric has the same form, with
$$
V=1-\frac{2GM}{r}+\frac{r^2}{b^2}
$$
It has a horizon at $r=r_+$, where $V(r_+)=0$. In both cases, the same change of coordinates as for Schwarzschild in flat space brings these 
metrics to unimodular form. We again go to Euclidean signature by taking $\tau=it$. Very near the horizon, a further change of coordinates
$$
r=r_++\sqrt{\frac{b^2+3r_+^2}{4b^2r_+^3}}y^2
$$
yields the metric
$$
ds^2\approx \frac{1}{r_+^2}\left(dy^2+y^2d(\frac{b^2+3r_+^2}{2b^2r_+}\tau)^2\right)+r_+^2d\bar \Omega_2^2
$$
which shows that the apparent singularity at $r=r_+$ can be removed if $\tau$ is regarded as an angular variable with period
$$
\beta=\frac{4\pi b^2 r_+}{b^2+3r_+^2}
$$
which corresponds to the same temperature as in GR (a further change from polar to Cartesian coordinates in the $\bR^2$ piece yields a unimodular metric).

Having revisited these solutions in unimodular form, we are now ready to evaluate the Euclidean action on them.  A straightforward evaluation would of course yield an infinite result, so we must regularize the action, by substracting a reference background. In ordinary GR, the reference backgrounds are taken to be the same metrics with $M=0$, i.e. empty Minkowski \cite{Gibbons:1976ue} or Anti de Sitter \cite{Hawking:1982dh} spaces, respectively. In particular the reference backgrounds have the same cosmological constant as the respective instantonic solutions. While this is a natural choice in GR, where the cosmological constant appears in the action, there is no such motivation in unimodular gravity, where $\Lambda$ is demoted to an integration constant. Instead, it is more natural to use as regulators backgrounds with the same coordinate 4-volume, even if they have different $\Lambda$ as the solution.

Since the evaluation of the action for the Schwarzschild solution in flat space turns out to be a particular case (setting $\Lambda=0$) of the Schwarzschild-Anti de Sitter case, we present the details for this  latter solution.

The physically relevant action is given by the difference of regularized 4-volumes for Schwarzschild Anti de Sitter and pure Anti de Sitter. However, as stressed, these two spacetimes need not have the same cosmological constant. Rather, we will fix the reference Anti de Sitter space by demanding that it has the same coordinate 4-volume 
as the Schwarzschild anti de Sitter solution. Demanding that the physical area of the $S^2$ at the cut-off is the same, we deduce that the cut-off $r=R$ must be the same for the two solutions.  Next we demand that the locally measured temperature is the same in both spacetimes (or in geometric language, that the physical length of the two $S^1$s is the same at $r=R$), which fixes the temperature of reference background (with cosmological constant $\Lambda'$) to be
$$
\beta' \sqrt{1+\frac{R^2}{b'^2}}=\beta \sqrt{1-\frac{2GM}{R}+\frac{R^2}{b^2}}
$$
where we left open the possibility that the two cosmological constants are different (in particular, even if $\Lambda=0$, it doesn't follow that $\Lambda'=0$). Finally, requiring that the two solutions have the same 4-volume implies
$$
\beta(R^3-r_+^3)=\beta' R^3
$$
Putting all this together, the difference in bulk actions is
$$
I_{bulk}=\frac{1}{16\pi G}\frac{4\pi}{3}\lim _{R\rightarrow \infty}\left(\beta(R^3-r_+^3)(4\Lambda-2L)-\beta'R^3(4\Lambda'-2L)\right)=\frac{\Lambda \beta}{3G}(r_+b^2-r_+^3)
$$
which is four times the GR result obtained by Hawking and Page \cite{Hawking:1982dh}. We are not done, however, since we still have to consider the boundary term,
$$
I_{bdy}=\frac{1}{8\pi G}\lim _{R\rightarrow \infty}\left(\int_{S-AdS}Kd\Sigma-\int_{AdS'}Kd\Sigma \right)
=-\frac{\Lambda \beta}{4G}(r_+b^2-r_+^3)
$$
which is -3 times the full GR result (in GR the boundary term vanishes for this computation  \cite{Hawking:1982dh, Witten:1998zw}). Adding both contributions we obtain,
$$
I_{bulk}+I_{bdy}=\frac{\Lambda \beta}{12G}(r_+b^2-r_+^3)
$$
which coincides with the GR result \cite{Hawking:1982dh}. This guarantees that using $I=-\hbox{log} Z$
one deduces the same energy and entropy for the solution as in GR. Although we presented the details for the Schwarzschild-Anti de Sitter solution ($\Lambda < 0$), by sending $\Lambda \rightarrow 0$ (and recalling $\Lambda b^2=-3$) we also recover the result of Gibbons and Hawking \cite{Gibbons:1976ue} for the action of the asymptotically flat Schwarzschild solution.\footnote{It is worth pointing out that in GR the result for $\Lambda=0$ comes exclusively from the boundary term, and for $\Lambda <0$ exclusively from the bulk term, while in the computation that we have presented here both terms contribute in all cases.} 

This computation can be repeated for arbitrary spacetime dimension $d$. We omit the details, indicating only the final result. We find that the bulk action is now $d$ times the GR result \cite{Witten:1998zw}, while the boundary term evaluates to $-(d-1)$ times the GR result, so after these two contributions are added, the full action evaluates to the same result as in GR, for arbitrary $d$.

\subsection{DeSitter}
We turn now to the case of $\Lambda >0$. The deSitter metric in global coordinates is ($\ell^2=3/\Lambda$),
$$
ds^2=-\ell^2d\tau^2+\ell^2\hbox{cosh}^2 \tau d\Omega_3^2
$$
It is easy to generalize Schwarzschild's change of coordinates to write a unimodular metric for $S^3$ (let $\theta_1$ be the polar angle in $S^3$ and introduce a variable $y$ with $4y=2\theta_1-\hbox{sin} (2\theta _1)$) and then a further change of coordinates
$$
T=\ell^4(\hbox{sinh }\tau+\frac{1}{3}\hbox{sinh}^3\tau)
$$
yields a unimodular metric for deSitter in global coordinates. We can pick up a Euclidean section corresponding to a 4-sphere of radius square $3/\Lambda$. 
All the contribution to the action comes from the bulk term, as there is no boundary
$$
I=-\frac{1}{16\pi G}(4\Lambda-2L)\frac{24 \pi^2}{\Lambda^2}=-\frac{3\pi}{G\Lambda}(2-\frac{L}{\Lambda})
$$
Using $I=\beta E-S$ and assigning $E=0$ to deSitter space, we conclude
$$
S=\frac{3\pi}{G\Lambda}(2-\frac{L}{\Lambda})=\frac{A}{4}(2-\frac{L}{\Lambda})
$$
where we didn't actually need to know the temperature of de Sitter  space\footnote{In \cite{Gibbons:1976ue}, the result given for this 
action and entropy is four times the standard result.}. Note that we recover the GR result only for $L=\Lambda$. If taken seriously, this 
is a disturbing result for various reasons. First, for $L>2\Lambda$, this entropy would be negative! Second, since it depends on $L$, 
it can't agree with any computation of the entropy using the metric (a la Hawking in his first paper), since $L$ does not appear in the metric. 

One might argue that the entropy is only determined up to an additive constant, and so the $L$ dependence might be spurious. Indeed,
in the present contex, we are considering the sum over histories with fixed four-volume, and therefore the $L$ term does 
not appear when we consider the change in the Euclidean action of two different solutions. We saw in the previous Subsection that this approach 
leads to sensible results in the limit of infinite volume. However, some puzzles remain in the case where the instantons have a finite volume.

In GR, for any given positive value of $\Lambda$, we have two different smooth Euclidean solutions: the de Sitter solution, which is a four-sphere,
and the Nariai solution, which is the direct product of two two-spheres of equal radii. In the Lorentzian continuation, the latter represents a
large black hole whose horizon is in thermal equilibrium with a cosmological horizon. The difference in the action 
of the two solutions is $\Delta I_{GR} = -\Delta A/4G$, where $\Delta A$ is the total change in horizon area between the two solutions.
In the unimodular theory, we can consider a de Sitter solution and a Nariai solution with the same four-volume. These will have different 
values for the integration constant $\Lambda\neq \Lambda'$, and different horizon areas, but the important point is that the difference in their 
action will be given by $\Delta I= -\Delta A/2G$. In other words, the relative entropy between two solutions is weighed by a factor of two 
with respect to the GR result. The significance of this result is unclear to us. Note that the Lorentzian versions of these solutions have infinite 
volume, and because of that 
it seems physically quite implausible that we may have a transition between two solutions with different values of the integration constant $\Lambda$. 
From this perspective, it is unclear why these two solutions should be compared to one another. On the other hand, we do not have 
any strong argument against the possibility that the area law may have a different coefficient in de Sitter space.

\subsection{Discussion}

We have just presented the evaluation of the action (\ref{unimod}) plus (\ref{ibdy}) for simplest Euclidean vacuum solutions, both for 
spaces with infinite and finite volume. For Schwarzschild black holes in asymptotic Anti de Sitter or Minkowski spaces, we have shown that, 
with the appropriate regularization prescription, we reproduce the GR result, in arbitrary spacetime dimension $d$.

In order to check if this agreement is peculiar to this particular solution or is more general, we can also consider solutions with matter. 
A simple example are Reissner-Nordstrom -Anti de Sitter black hole solutions of Einstein-Maxwell theory in arbitrary dimension, and with a 
negative cosmological constant. The thermodynamics of these solutions has been discussed in \cite{Chamblin:1999hg}, both at fixed charge 
and at fixed potential. We have considered the case of fixed potential, and again find complete agreement for the regularized actions of 
the unimodular theory and General Relativity, in arbitrary spacetime dimension. These examples, together with the perturbative argument of 
section 2, point towards the possibility that this formulation of unimodular gravity is semiclassically equivalent to GR in spacetimes with 
infinite volume. It would be important to elucidate if this agreement extends to arbitrary solutions with matter.

On the other hand, for the simplest solution with compact volume, i.e. de Sitter, the result we find not only disagrees with the one from GR, 
but is also hard to interpret. Therefore, it is currently unclear to us that the action 
(\ref{unimod}) plus (\ref{ibdy}) can be a valid starting point to quantize unimodular gravity in spacetimes with finite 4-volume. This issue 
deserves further investigation.

\section{Generalized unimodular gravity}

The discussion in Sections 2 and 3 suggests that we can generalize unimodular gravity to a theory where we integrate
over all metrics of constant determinant $g=a^2$ , where the constant $a$ is integrated over. This is the same as allowing 
for different coordinate 4-volumes in the sum over histories with the unimodular action (\ref{unimod}), since
a change in the coordinate 4-volume can always be reabsorbed by a constant rescaling of the metric.
Interestingly,  the addition of this single global degree of freedom is suficient in order to make the theory classically 
equivalent to GR. As we shall see, there is freedom in this generalization and the different choices need not be equivalent quantum mechanically.

The generating functional defining this new theory is given by 
\begin{equation} 
Z[J] \propto \int_{-\infty}^\infty w(a)da \int D\hat g_{\mu\nu} D\psi \ 
e^{iI[a^{1/2}\hat g_{\mu\nu},\psi,J]}, \label{pgf3} 
\end{equation} 
where by definition $\hat g_{\mu\nu}$ has unit determinant 
\footnote{Here we depart from the notation used in the rest of the paper, 
where the metric of unit determinant is simply refered to as $g_{\mu\nu}$.}
, and $D\hat g_{\mu\nu}$ is invariant under transverse 
diffeomorphisms. The weight $w(a)$ is arbitrary, so there is  freedom in defining the
theory. If $w(a)=\delta(a-1)$, then we recover the standard unimodular gravity. However, if $w(a)$ is a smooth
function, then we recover GR in the classical limit.

Indeed, it is straightforward to check that the full set of Einstein's 
equations (and not just the traceless part) can be recovered from the action 
when variation with respect to $a$ is included.  
First of all, variation with respect to $\hat g_{\mu\nu}$ yields the traceless part of Einstein's equations. 
Second, the matter equations of motion yield the conservation of the matter stress tensor. Using this and the 
Bianchi identity for $G_{\mu\nu}$ in 
the divergence of the traceless part of Einstein's equations, one gets the result that the trace of Einstein's equations is equal 
to an arbitrary integration constant $C$, just as in unimodular gravity. However, the derivative with respect to $a$ tells us that 
$\int d^4 x\ C  =0$, and so $C=0$. Here, we are neglecting the (imaginary part) of $d\ln w(a)/da$ in comparison to $(1/\hbar) dI/da$,
where we have reintroduced Planck's constant. Hence, the full set of Einstein's equations is recovered in the classical limit 
(in a gauge in which the determinant of the metric is constant). 

Three comments are in order:

\begin{itemize}

\item First, since now all of Einstein's equations are recovered, $\Lambda=L$ is no longer
and ad-hoc imposition, and so the thermodynamical arguments based on the semiclassical analysis of the 
Euclidean action will not lead to any problems: the same relations as in GR will be recovered. 

\item Second, the unimodular case $w(a)=\delta(a-1)$ is singular, in the sense that derivatives of the weight function
cannot be neglected even in the semiclassical limit. Because of that, the trace of Einstein's equation 
is not recovered. 

\item And third, since the weight function $w(a)$ is arbitrary, the quantum theory may depend the particular choice of $w(a)$ 
and may thus be different from GR.

\end{itemize}

\section{Generally covariant version of unimodular gravity}

In Ref. \cite{Henneaux}, Henneaux and Teitelboim studied unimodular gravity in the Hamiltonian formalism. They
pointed out that the number of constraints is larger than suggested by the naive counting associated with the 
invariance under transverse diffeomorphisms. Technically, this is due to a ``tertiary'' constraint 
which arises when we require that the standard ``momentum constraint" associated with spatial diffeomorphisms 
be preserved in time. Because of that, the constraints are almost the same as in GR, 
except for the fact that the theory is not invariant under reparametrizations of the time variable $T \to f(T)$.
Henneaux and Teitelboim then suggest a method for making the theory invariant under time reparametrizations 
without changing its dynamical content. After going back to the Lagrangian formalism, the action takes the 
generally covariant form
\begin{equation}
I=\int [(-g)^{1/2} (R-2\Lambda) + 2\Lambda \partial_\mu \tau^\mu] d^4 x. \label{ht0}
\end{equation}
Here, $\tau^{\mu}$ is a vector density, and the action has to be extremized with respect to 
$g_{\mu\nu}(x)$, $\Lambda(x)$ and $\tau^\mu(x)$. This leads to the equations of motion 
\begin{eqnarray}
R_{\mu\nu} -{1\over 2} R g_{\mu\nu} + \Lambda g_{\mu\nu} &=&0,\\
\Lambda,_\mu &=& 0,\\
\partial_{\mu}\tau^\mu -(-g)^{1/2}&=& 0. \label{ht3}
\end{eqnarray} 
The first two equations coincide with those of unimodular gravity in the gauge $g=-1$, which we can always 
adopt by a suitable choice of the vector density $\tau^{\mu}$ [see Eq. (\ref{ht3})].

Let us now discuss semiclassical gravity. If we were to take the action (\ref{ht0}) (with the addition of the
Gibbons-Hawking boundary term) as the relevant one to be used in the semiclassical evaluation of the partition 
function, then it is clear that on-shell, the second and third terms would cancel 
each other, and the action would reduce to the first term, just as in the case of unimodular gravity. 
Hence, we would find the same results which we have discussed in the previous 
subsections, which disagree with those of GR. 

However, we note that to the action (\ref{ht0}) we may add a total derivative of the form
\begin{equation}
\Delta I= C \int \partial_{\mu} (\Lambda \tau^{\mu}),
\end{equation}
where $C$ is an arbitrary constant. This will not change the equations of motion, but it will modify the 
value of the on-shell action. The ambiguity in $C$ can be resolved by demanding that the action should be 
stationary under variations $\delta \Lambda$ that vanish at the boundary, while variations $\delta \tau^\mu$
are not required to vanish. This corresponds to the choice $C=-2$. The choice is motivated by the fact 
that $\Lambda$ is physically measurable (since it determines the curvature) while $\tau^\mu$ is not. 
In fact, the action is invariant under
gauge transformations of the form $\delta \tau^\mu = \epsilon^\mu$, where $\epsilon^\mu$ is an arbitrary vector 
density satisfying $\partial_\mu\epsilon^\mu=0$ \cite{Henneaux}.
\footnote{A similar ambiguity arises in the theory of gravity coupled to a three form $A$ with field strength 
$F=dA$, and kinetic term in the action proportional to $F^2$
(See e.g. the discussion in \cite{BP}, and references therein). In four dimensions, this field is not dynamical, 
and the value of the four form is given by an integration constant. Then, the $F^2$ term in the action leads to 
a term in the equations of motion which mimicks a cosmological constant. This is similar to the situation in the 
theory (\ref{ht0}) except that the $F^2$ term has definite sign, while $\Lambda$ in  (\ref{ht0}) 
can have either sign.  In this theory, 
the boundary term for the three form is chosen so that the equations of motion 
follow from variations such that $\delta F=0$ at the boundary, while $\delta A$ is unrestricted.}
 Including all boundary terms, the action takes the form
\begin{equation}
I=\frac{1}{16\pi G}\int [(-g)^{1/2} (R-2\Lambda) - 2\Lambda,_\mu \tau^\mu] d^4 x 
+ \frac{1}{8\pi G}\int_{\partial Y}d^3x \; \sqrt{-h}K.
\end{equation}
It is clear that on-shell, the term proportional to $\Lambda,_\mu$ will vanish, leading to the standard 
semiclassical results of GR.

\section{Acknowledgements} 
We would like to thank Enrique {\'A}lvarez, Diego Blas, 
Marco Caldarelli, Guillem P{\'e}rez, Takahiro Tanaka and Paul Townsend for discussions. 
This work was supported in part by by grants FPA2007-66665C02-02 and  
DURSI 2005-SGR-00082. B.F. is also supported by a Ram{\'o}n y Cajal fellowship.

\end{document}